# Emergent magnetism, heavy electrons and pressure-induced reentrant superconductivity in the iron substituted 4*d* transition-metal sulfides

Yalei Huang[1, 4, 15], Lu Xin[2, 15], Na Zuo[3, 15], Bin Li[5,6], Wei Zhou[7], Dhanarajagopal Alltrin[1], Boning Yu[1], Haiyang Yang[8], Bin Qian[7], Wen-Chin Lin[9], Raman Sankar[10], Michael Smidman[11], Xiangzhuo Xing[3, *], Chunqiang Xu[12, *], Xiaobing Liu[3, *], Jianhui Dai[13], Dong Qian[14], Shiyan Li[2, *] and Xiaofeng Xu[1, 4, *]

[1]*Zhejiang Provincial Key Laboratory of Quantum Precision Measurement, School of Physics and Optical Engineering, Zhejiang University of Technology, Hangzhou 310023, China*

[2]*State Key Laboratory of Surface Physics, Department of Physics, Fudan University, Shanghai 200438, China*

[3]*Laboratory of High Pressure Physics and Material Science (HPPMS), School of Physics and Physical Engineering, Qufu Normal University, Qufu 273165, China*

[4]*ZJUT Yinhu Research Institute of Innovation and Entrepreneurship, Fuyang District, Hangzhou 311400, China*

[5]*School of science, Nanjing University of Posts and Telecommunications, Nanjing 210023, China*

[6]*Jiangsu Physical Science Research Center, Nanjing 210093, China*

[7]*School of Electronic and Information Engineering, Suzhou University of Technology, Changshu 215500, China*

[8]*College of Materials and Environmental Engineering, Hangzhou Dianzi University, Hangzhou, China*

[9]*Department of Physics, National Taiwan Normal University, Taipei 116, Taiwan*

[10]*Institute of Physics, Academia Sinica, Nankang, Taipei, R.O.C. 11529, Taiwan*

[11]*School of Physics, Zhejiang University, Hangzhou, China*

[12]*School of Physical Science and Technology, Ningbo University, Ningbo 315211, China*

[13]*School of Physics, Hangzhou Normal University, Hangzhou 310036, China*

[14]*Key Laboratory of Artificial Structures and Quantum Control (Ministry of Education), School of Physics and Astronomy, Shanghai Jiao Tong University, Shanghai 200240, China*

[15]These authors contributed equally: Yalei Huang, Lu Xin, Na Zuo.

[*]Corresponding authors. Email: xzxing@qfnu.edu.cn, xuchunqiang@nbu.edu.cn, xiaobing.phy@qfnu.edu.cn, shiyan_li@fudan.edu.cn, xuxiaofeng@zjut.edu.cn

## Abstract

Superconductivity emerging from or in the vicinity of magnetic states is generally considered to be mediated by spin fluctuations and thus lies beyond the scope of conventional electron-phonon coupled BCS framework. Here we report the emergence of novel ferromagnetism in the *d*-electron rhodium sulfide $Rh_{17}S_{15}$ superconductor, characterized by an enhanced Sommerfeld coefficient $\gamma$ arising from the flat topological band and many-body correlations. We further demonstrate that the ferromagnetism can be tuned via Fe substitution at the Rh sites, leading to a spin glass ground state induced by the competing ferromagnetic and antiferromagnetic exchange interactions. Fe doping results in a further enhancement of both the $\gamma$ and electron effective masses. At a doping level of $x = 0.67$ in $Rh_{17-x}Fe_xS_{15}$, $\gamma$ reaches 312 mJ mol$^{-1}$K$^{-2}$, second only to the well-documented *d*-electron heavy-fermion material $LiV_2O_4$. Furthermore, upon applying pressure, superconductivity is first suppressed; under high pressures, however, we observe the reentrant superconductivity in both pristine and Fe-doped samples. Our results not only demonstrate the unusual magnetic states and possible heavy-fermion features in these frustration-free, *d*-based superconductors, but also suggest that the superconductivity in this system is likely mediated by the intrinsic spin fluctuations and may thus be *unconventional*.

## Introduction

Unconventional superconductivity typically emerges either directly from a magnetic normal state or at the boundary of a magnetic phase, where its occurrence is tuned via external parameters such as pressure, magnetic field, or chemical doping [1-3]. These unconventional superconductors, epitomized by high-$T_c$ cuprates, exhibit a wide range of exceptional characteristics, such as a non-*s*-wave order parameter and a strictly *T*-linear resistivity that extends over a decade in temperature *T*, a phenomenon dubbed strange metallicity [4]. Another broad family of widely studied unconventional superconductors are heavy-fermion superconductors, featuring vastly enhanced electron effective masses by tens to thousands of times the bare electron mass [3]. The origin of heavy-fermion behavior is generally attributed to the Kondo interaction between localized moments and the conduction electron Fermi sea [5]. Notably, the majority of these heavy-fermion superconductors are found in *f*-electron based compounds, due to the strongly compact nature of *f*-orbitals relative to the atomic nucleus [6]. Contrarily, heavy-fermions in *d*-based compounds are extremely rare and reported only in a handful of *d*-electron materials [7-9]. One of the most outstanding examples of *d*-based heavy fermions is the heavy-fermion state in the spinel-type transition-metal oxide $LiV_2O_4$, whose origin remains debated although magnetic frustration is generally believed to play a pivotal role in its formation [10-13]. Recent reports of heavy *d*-electron compounds include the kagome metal $Ni_3In$ and the three-dimensional flat-band pyrochlore $CaNi_2$, although their microscopic mechanisms underlying the mass enhancement may differ markedly and remain the subject of active debate [14,15]. From an electronic band structure perspective, all these heavy-fermion compounds are characterized by flat bands with negligible dispersions along certain or all spatial dimensions, a hallmark of strong electron correlations that give rise to a wide spectrum of novel quantum states and phenomena [16].

Among unconventional superconducting materials, nearly all are artificially synthesized and do not occur naturally. The possible exception is the rhodium sulfide $Rh_{17}S_{15}$, known as miassite, which exists in nature as a mineral in the placers along the Miass river and thus got its name after that [17]. Remarkably, this material clinches the highest superconducting $T_c$ (= 5.0 K) among all naturally occurring superconductors discovered to date. Besides, this superconductor manifests numerous unconventional characteristics in both the normal and superconducting states. For example, unlike conventional non-interacting or weakly correlated metals, the resistivity of $Rh_{17}S_{15}$ exhibits a broad hump around 60 K, across which the Hall coefficient reverses its sign; this behavior is reminiscent of strongly correlated electron systems [18,19]. Upon further cooling, a superconducting transition set in below $T_c$ = 5.0 K [20-22]; this value may vary slightly with sample stoichiometry, owing to the volatility of sulfur during crystal growth. More intriguingly, the Sommerfeld coefficient derived from heat capacity measurements reaches a magnitude of $\gamma$ = 105 mJ/mol K$^2$, significantly higher than that of weakly correlated metals and 3-5 times larger than that predicted from the band structure calculations [22-24], suggesting the strong electron correlations in this system. Moreover, recent susceptibility measurements have revealed a robust weak ferromagnetism in its normal state [19], leading to the conjecture that superconductivity is mediated by ferromagnetic spin fluctuations and that its pairing mechanism may also be unconventional. Additionally, its superconductivity features a large heat capacity jump at $T_c$ ($\frac{\Delta C}{\gamma T_c}$ = 2, substantially larger than the BCS value of 1.43) and a large upper critical field $H_{c2}$ that exceeds the Pauli paramagnetic limit by a factor of 2 [18,22]. Lastly, our recent pressure measurements have uncovered a second superconducting dome at high pressures [19]. Note that the reemergent superconductivity under high pressure is only observed in a small subset of unconventional superconductors [25-27]. All these remarkable features underscore the highly unusual physical properties of $Rh_{17}S_{15}$, thereby enshrining it in the family of putative unconventional superconductors.

What is not evident from previous studies is whether the weak ferromagnetism and superconductivity observed in pristine $Rh_{17}S_{15}$ retain their robustness when the system is intentionally doped with certain (ferro)magnetic elements. This doping strategy may further prove or disprove the claim that the ferromagnetism reported in earlier work stems from contamination by ferromagnetic impurities. The other outstanding open question is whether a second superconducting dome can also be induced in these magnetically doped samples under applied pressure.

In this work, we introduced a small amount of magnetic Fe into the $Rh_{17}S_{15}$ system via chemical substitution, i.e., $Rh_{17-x}Fe_xS_{15}$, and systematically investigated the evolution of the thermodynamic and transport properties as a function of doping level. Notably, the weak ferromagnetism observed in the pristine $Rh_{17}S_{15}$ is replaced by spin glass states in the Fe-doped crystals, due to the competition between ferromagnetic and antiferromagnetic interactions. Meanwhile, Fe doping results in a further enhancement of both $\gamma$ and electron effective masses. Though ambient-pressure superconductivity is progressively suppressed with Fe doping, we observed robust reentrant superconductivity under high pressures, with superconducting $T_c$ exceeding that at the ambient pressure. All these findings collectively point towards the emergence of exceptional superconducting states, both at ambient pressure and under high pressure, which may be closely associated with spin fluctuations present in this $Rh_{17-x}Fe_xS_{15}$ system.

## Results

### Crystal structure and electronic structure

$Rh_{17}S_{15}$ crystallizes in a cubic structure with the space group $Pm\overline{3}m$ [18,28]. As illustrated in Fig. 1a and Fig. S1, Rh atoms occupy four symmetry-inequivalent sites (1*b*, 24*m*, 3*d*, and 6*e*) and S atoms reside at three sites (12*i*, 12 *j*, and 6 *f*). Specifically, the Rh 1*b* site is situated at the body-centered position, with the Rh 24*m* site forming a cage around it. The Rh 3*d* site resides in the middle of two Rh 6*e* sites on the edge of the unit cell. First-principles calculations were carried out to investigate the electronic structure of $Rh_{17}S_{15}$ and the results demonstrate that this compound exhibits the characteristic features of an enforced semimetal with nontrivial topological properties (Fig. 1b-d). Two bands cross the Fermi level ($E_F$) with pronounced degeneracy near the R point along the Γ-R-X path, while the band structure shows mirror symmetry with respect to $E_F$, particularly around the Γ point (Fig. 1b). Flat band regions near $E_F$, highlighted along the R-X-M-R path, give rise to sharp van Hove singularities (vHs) at approximately +0.01 eV and -0.03 eV in the density of states, suggesting enhanced electronic correlation effects (Fig. 1c). Atom-resolved analysis reveals that Rh atoms occupying the 24m Wyckoff position dominate the electronic states near the Fermi level, accounting for over 60% of the total DOS at $E_F$. The electronic specific heat coefficient is estimated to be $\gamma$ = 35.2 mJ mol$^{-1}$ K$^{-2}$. As will be shown later, the experimental specific coefficient is 3 times larger than the calculated value, indicating strong electron correlations in this material. Surface state calculations for the (001) termination reveal multiple non-trivial topological surface bands with dispersions that traverse the Brillouin zone edge (Fig. 1d). The $Z_2$ topological indices ($\nu_0$; $\nu_1\nu_2\nu_3$) were determined by computing the evolution of Wannier charge centers (WCCs) across the six time-reversal invariant momentum planes at $k_x = 0, \pi$, $k_y = 0, \pi$, and $k_z = 0, \pi$. Our calculations reveal that the system is in a strong topological phase characterized by $Z_2$ indices (1; 011), where $\nu_0 = 1$ indicates nontrivial bulk topology. The weak indices show $\nu_1 = 0$ for the $k_x = \pi$ plane, while $\nu_2 = \nu_3 = 1$ for the $k_y = \pi$ and $k_z = \pi$ planes, confirming the topologically nontrivial nature of the electronic structure. Symmetry-based topological classification through irreducible representations at high-symmetry $k$-vectors, combined with compatibility relations and elementary band representation (EBR) theory [29], conclusively identifies $Rh_{17}S_{15}$ as an enforced semimetal wherein the Fermi-level band crossings are symmetry-protected and cannot be gapped without breaking crystalline symmetries, positioning $Rh_{17}S_{15}$ as a compelling platform for exploring topological quantum phenomena in correlated electron systems.

The $Rh_{17-x}Fe_xS_{15}$ single crystals investigated in this work were synthesized by the self-flux method (details given in the Experimental Section in Supporting Information). Their chemical compositions were determined by energy-dispersive X-ray spectroscopy (EDX), with the Fe doping level ranging from $x = 0$ to $x = 0.67$. The doping levels were cross-verified with the electron probe microanalysis (EPMA) across different regions of the crystals. The results show a consistent doping level with EDX and homogeneous elemental distributions throughout the samples. It is worth noting that, as Rh occupies four distinct crystallographic sites, the specific Rh site substituted by Fe cannot be unambiguously determined in this study. As shown in Fig. 1e, EDX elemental mapping confirms the homogeneous spatial distribution of Rh, Fe and S elements across the as-grown single crystals, verifying the absence of elemental segregation during synthesis. Fig. 1f displays the X-ray photoelectron spectroscopy (XPS) spectra of the Rh 3*d* and S 2*p* regions for the pristine $Rh_{17}S_{15}$ samples, both of which are well-fitted by the simulated curves. Fig. 1g and Fig. S2 show the powder X-ray diffraction (XRD) for $Rh_{17}S_{15}$ and the highest Fe-doped sample ($x$ = 0.67). No additional diffraction peaks are detected, and all observed reflections can be indexed to the $Rh_{17}S_{15}$-type structure (space group $Pm\bar{3}m$, PDF #73-1443). The diffraction peaks of $Rh_{17-x}Fe_xS_{15}$ ($x$ = 0.67) are marginally shifted to lower angles compared with $Rh_{17}S_{15}$, indicating a slight increase in the lattice parameter upon Fe doping. These results indicate that the crystal structure of $Rh_{17-x}Fe_xS_{15}$ is essentially retained upon Fe substitution, with no impurity phases formed.

**Anomalous transport and enhanced effective masses**

To investigate the effect of Fe substitution on the superconducting and transport properties of $Rh_{17-x}Fe_xS_{15}$, zero-field temperature dependence of resistivity ($\rho$-$T$) measurements were performed for all samples, as encapsulated in Fig. 2a. The low-temperature (low-$T$) resistance for $Rh_{17-x}Fe_xS_{15}$ ($x$ = 0.27, 0.34, 0.48, and 0.67) were measured in a $^3$He cryostat (Fig. S7, S10, S12 and S14). The superconducting transition temperature ($T_c$), defined as the onset temperature of superconductivity, is progressively suppressed by Fe doping, decreasing from 5.3 K for ($x$ = 0) to 1.85 K for ($x$ = 0.48). . For $x$ = 0.67, no superconductivity is observed down to 0.4 K. Reminiscent of many correlated compounds, the pristine $Rh_{17}S_{15}$ and $Rh_{17-x}Fe_xS_{15}$ ($x$ = 0.10) displays a broad resistivity hump near 60 K upon cooling (Fig. S4 and Fig. S6). Furthermore, the magnetic-field dependence of Hall resistivity ($\rho_{yx}$-$\mu_0H$) was measured at various temperatures, as shown in Fig. 2b. As seen, the $\rho_{yx}$ curves at 10 K for all samples exhibit a negative slope and are quasi-linear in field, indicating dominant electron-type carriers at low temperatures. The electron carrier density $n_e$ can thus be estimated from $n_e = 1/(eR_H)$, where $R_H$ is the Hall coefficient extracted from the linear fits to the $\rho_{yx}$-$\mu_0H$ curves. The resulting phase diagram is illustrated in Fig. 2c. With increasing Fe content ($x$), superconductivity is gradually suppressed, whereas the carrier density $n_e$ initially increases with doping and then plateaus upon further Fe substitution.

Interestingly, a resistivity upturn appears in the $\rho$-$T$ curves of the doped samples ($x$ = 0.27 and $x$ = 0.34), with minima observed at 33 K and 35 K, respectively (Fig. 2a). As $x$ increases further, this upturn disappears for $x$ = 0.48 and $x$ = 0.67, where the resistivity exhibits a nearly linear-in-$T$ dependence at low temperatures (Fig. S12 and S14). We now focus on the resistivity upturn behavior. Taking the $x$ = 0.27 sample as an example, Fig. S7 presents the low-$T$ resistivity plot on a semi-logarithmic scale, revealing a clear-cut logarithmic increase in resistivity with decreasing temperature, which is consistent with a Kondo-like scattering scenario [30-33]. Moreover, this resistivity upturn is suppressed by the application of a magnetic field (Fig. S7), reflecting the splitting of the Kondo resonance induced by magnetic fields [34,35]. It is worth noting that electron-electron interaction corrections and weak localization may also give rise to low-$T$ resistivity upturns. However, these mechanisms are unlikely to be the dominant origin for the present upturn behavior because they are not consistent with our temperature-dependent transport data. Further details and discussion are provided in the Supplementary Information.

Furthermore, Magnetoresistance (MR) measurements provide additional support for this interpretation. Fig. 2d shows the MR curves of $Rh_{17-x}Fe_xS_{15}$ at 10 K. For $x = 0$ and 0.1, positive MR is observed with a quadratic dependence on the applied magnetic field. As the Fe concentration increases, samples with $x = 0.27$, 0.34, and 0.67 display negative MR at low fields, which crosses over to positive MR at higher fields. Similar negative MR behavior has been reported in Kondo systems [36,37]. These results suggest that the resistivity minimum at 33 K mark the onset of Kondo scattering [38], denoted as $T_{\mathrm{K}}^{\mathrm{onset}}$. With further decreasing temperature, the resistivity of the $x = 0.27$ sample decreases again below $T = 7$ K (Fig. 2e). This behavior contrasts sharply with that of a system with independent Kondo impurities, where the resistivity saturates at low temperatures [39-42]. Instead, the resistivity of the $x =$ 0.27 sample bears a striking resemblance to that of the heavy-fermion compounds such as $CeCoIn_5$ [43] and $CeCu_6$ [44], which show a logarithmic rise below $T_{\mathrm{K}}^{\mathrm{onset}}$ and a peak at the coherence temperature $T^*$. Below $T^*$, as coherence develops among the scatterers, the resistivity begins to decrease again (Fig. S3). We therefore identify $T^* = 7$ K as the coherence temperature for $x = 0.27$ sample, which is substantially lower than $T^* = 60$ K observed in the parent compound (Fig. S4). The characteristic temperature scales for $x = 0.27$ are marked in Fig. 2e, along with schematic diagrams of the Kondo-lattice screening temperature regimes in Fig. 2f. Hybridization between localized electrons and conduction electrons gives rise to Kondo screening [32]. This screening first emerges in a local and incoherent form below $T_{\mathrm{K}}^{\mathrm{onset}}$. Upon further cooling to $T < T^*$, the Kondo screening becomes coherent, manifested by a renormalized Kondo resonance, and the ground state exhibits a large quasiparticle effective mass.

To further explore the Kondo-like behavior, the zero-field heat capacity was measured for all $Rh_{17-x}Fe_xS_{15}$ samples at low temperatures, as shown in Fig. 2g. A characteristic jump associated with the superconducting transition at $T_c$ is observed for $x = 0$ and $x = 0.1$, indicative of a bulk thermodynamic transition. With increasing Fe content, this jump vanishes within the measured temperature range (down to 2 K). The total heat capacity can be described by the expression,

$$C = \gamma T + \beta T^3 + \delta T^5 \tag{1}$$

where $\gamma T$ represents the electronic contribution and $\beta T^3 + \delta T^5$ accounts for the lattice contribution. The fitting yields $\gamma$ = 113 mJ mol$^{-1}$ K$^{-2}$ for $Rh_{17}S_{15}$, consistent with previously reported values [18,45]. For $x = 0.27$, a moderately high $\gamma$ of 200 mJ mol$^{-1}$ K$^{-2}$ is obtained. This value is remarkably large among *d*-electron based materials and comparable to that of *f*-electron heavy-fermion materials [7-9,42]. Based on the derived carrier concentration and the heat capacity $\gamma$ coefficient, the electron effective mass $m^*$ can be estimated [19,24,45]. For $Rh_{17}S_{15}$, the calculated $m^*$ reaches 28 $m_0$ ($m_0$ is the free electron mass). The extracted values of $\gamma$ and $m^*$ for $Rh_{17-x}Fe_xS_{15}$ are summarized in Fig. 2h. As shown, both $\gamma$ and $m^*$ increase gradually with Fe doping. At $x = 0.67$, $m^*$ is enhanced to 45 $m_0$ and $\gamma$ reaches a maximum value of 312 mJ mol$^{-1}$ K$^{-2}$, nearly three times larger than that of the parent compound. This value is only preceded by $LiV_2O_4$ ($\gamma$ = 420 mJ mol$^{-1}$K$^{-2}$) [46] among all *d*-electron heavy fermions.

From the fitted $\beta$ values, the Debye temperature $\Theta_D$ can be estimated using the relation,

$$\Theta_D = \sqrt[3]{12\pi^4 N r k_B / 5\beta} \tag{2}$$

where $N$ is the Avogadro's number, $r$ is the number of atoms per formula unit and $k_B$ the Boltzmann constant. For $x = 0$, the estimated $\Theta_D$ is 472 K. As shown in Fig. 2i, $\Theta_D$ remains nearly constant up to $x = 0.34$ and then decreases abruptly to 420 K at $x = 0.67$. Moreover, the electron-phonon coupling parameter $\lambda_{e\text{-}p}$ was calculated using the McMillan's theory [47],

$$\lambda_{e-p} = [1.04 + \mu^* \ln(\Theta_D/1.45T_c)]/[(1 - 0.62\mu^*)\ln(\Theta_D/1.45T_c) - 1.04] \tag{3}$$

where the Coulomb pseudopotential $\mu^*$ is assumed to be 0.13. The calculated $\lambda_{e\text{-}p}$ values for different Fe concentrations are shown in Fig. 2i. As $x$ increases, $\lambda_{e\text{-}p}$ decreases gradually, mirroring the suppression trend of $T_c$ (Fig. 2c). On the other hand, for three-dimensional systems, the Fermi temperature $T_F$ is given [48,49],

$$T_F = \tfrac{1}{2}\hbar^2 (3\pi^2)^{\frac{2}{3}} \left[ n^{\frac{2}{3}} / k_B m^* \right] \tag{4}$$

where $\hbar$ is the reduced Planck constant, $n$ is the carrier concentration and $m^*$ is the electron effective mass. Using the values for $Rh_{17-x}Fe_xS_{15}$, we constructed the Uemura plot ($T_c$ vs. $T_F$), as shown in Fig. 2j. Notably, it is found that $Rh_{17-x}Fe_xS_{15}$ samples with $x \leq 0.67$ fall well within the unconventional superconductivity regime, suggesting that the superconductivity in this system may be unconventional.

**Unusual weak ferromagnetism and spin-glass state**

To investigate the evolution of magnetic properties in $Rh_{17-x}Fe_xS_{15}$, temperature-dependent magnetization (*M-T*) measurements were performed in both zero-field-cooled (ZFC) and field-cooled (FC) modes, as shown in Fig. 3a. With increasing $x$, superconductivity is suppressed, consistent with the electrical transport results. Notably, for $x \geq 0.27$, the ZFC and FC branches bifurcate at a finite temperature, and a peak emerges at $T_f$ in the ZFC curves, which is a typical signature of a spin-glass transition. To further clarify this behavior, Fig. 3b presents the temperature dependence of the in-phase component of *ac* susceptibility ($\chi'$-$T$) for $x = 0.67$ under different frequencies. The $\chi'$-$T$ curves exhibit a distinct frequency dependence. As the frequency increases, the peak shifts toward higher temperatures and its magnitude decreases. This is a hallmark feature of spin glass states [50]. Isothermal magnetization (*M-H*) curves for different Fe doping are shown in Fig. 3c, Fig. S8, S11, S13 and S15. For $x \geq 0.27$, all curves exhibit hysteresis. For $x \leq 0.48$, the coercive field ($H_c$) remains of the order of several hundred Oe and increases slightly with $x$, whereas it reaches 1800 Oe for $x = 0.67$ (Fig. 3c). These results are summarized in the temperature-composition phase diagram presented in Fig. 3d.

Moreover, weak ferromagnetism (WFM) has previously been observed in the parent compound $Rh_{17}S_{15}$, persisting from low temperatures up to room temperature, likely originating from local magnetic moments of Rh ions (see Fig. S5) [19]. To examine whether this intrinsic WFM persists with Fe doping, *M-H* curves at 300 K were measured for $Rh_{17-x}Fe_xS_{15}$, as shown in Fig. 3e and 3f. A clear hysteresis loop is seen for $Rh_{17}S_{15}$ at 300 K, with $H_c = 75$ Oe (Fig. 3e and Fig. S5). The presence of WFM is further corroborated by the magneto-optical Kerr effect (MOKE) measurements (Fig. 3g), where the Kerr rotation angle exhibits distinct hysteresis loops at 100, 200, and 300 K. With increasing $x$, the room-temperature WFM gradually weakens and eventually disappears for $x = 0.67$, where the *M-H* curve becomes linear, typical of a paramagnetic response (Fig. 3f). This observation also confirms that the WFM in $Rh_{17}S_{15}$ is intrinsic rather than arising from magnetic impurities. Overall, these results demonstrate that Fe doping suppresses superconductivity and induces a low-$T$ spin-glass state, while the room-temperature WFM progressively weakens and vanishes at high Fe concentrations in $Rh_{17-x}Fe_xS_{15}$.

**Pressure-induced reentrant superconductivity**

The novel magnetic phenomena uncovered in the normal states of $Rh_{17-x}Fe_xS_{15}$ suggest that their ambient-pressure superconductivity may be unconventional. As a well-established thermodynamic tuning parameter, pressure provides a clean and powerful tool for exploring superconducting physics [27,51-55]. To explore this further, we employed the diamond-anvil-cell technique (DAC) to apply high pressures for electronic transport measurements (Fig. 4a). Fig. 4b illustrates the renormalized resistance $R/R$(300 K) of $Rh_{17-x}Fe_xS_{15}$ ($x$ = 0.27) under various pressures. Within the studied pressure range, all resistivity curves exhibit metallic behaviors, i.e., resistivity decreases with lowering temperature (Fig. S9). Superconductivity is rapidly suppressed by pressure initially, accompanied by an initial increase in overall resistance. Above 80 GPa, resistivity drops sharply and superconductivity reemerges, with $T_c$ significantly higher than the ambient-pressure value. This result is akin to the behavior of the parent $Rh_{17}S_{15}$ under high pressure[19], where a structural transition as the origin of high-pressure superconducting phase can be ruled out by synchrotron XRD measurements (Fig. S17). These results are summarized in the phase diagram presented in Fig. 4c.

We further studied the most Fe-doped sample ($x$ = 0.67) under high pressure. Fig. 4d incorporates all resistivity curves measured up to 130 GPa, revealing distinct changes in the resistivity profile across two characteristic pressures, $P_1$ and $P_2$. Below $P_1$, the curves exhibit metallic behavior, that is, resistivity decreases with decreasing temperature. Between $P_1$ and $P_2$, the curves switch to insulating behavior, i.e., it increases with decreasing temperature. Above $P_2$, metallic behavior is restored. These crossovers in $R(T)$ profiles are more clearly illustrated in Fig. 4e. Interestingly, despite the absence of superconductivity down to 0.4 K at ambient pressure for this doping, a superconducting transition emerges when $P \geq P_2$. As depicted in Fig. 4f, superconductivity starts to develop at 85.57 GPa. At higher pressures, zero-resistance superconductivity is observed, with the superconducting $T_c$ continuously rising to 4.5 K with pressure up to 130.90 GPa.

The bulk nature of the observed superconductivity for $x$ = 0.67 was confirmed by its upper critical field and the characteristic $I$-$V$ curves at 121.92 GPa, presented in Fig. 4g and 4h, respectively. $T_c$ at different fields was measured by sweeping temperature at a constant field, as plotted in Fig. 4g for $P$ = 121.92 GPa. The $T_c$ values, determined by the onset of the transition at various fields, are summarized in the inset of Fig. 4g. We fitted the $\mu_0 H_{c2}$ curve using the Ginzburg-Landau (GL) theory, yielding the upper critical field at zero temperature $\mu_0 H_{c2}(T = 0\ \text{K})$ = 9.9 T. This value is marginally higher than the weak-coupling Pauli paramagnetic limit $1.84T_c$ (= 8.3 T), signifying unusual superconductivity. Fig. 4h shows the $I$-$V$ curves measured at several temperatures below $T_c$ under 121.92 GPa, which are consistent with the bulk superconductivity at such high pressures. Furthermore, MR and the Hall resistance measurements were taken at 10 K under various pressures, as shown in Fig. 4i and Fig. S16, respectively. The pressure dependences of the Hall resistance slope $\alpha$ and the MR value at 5 T are displayed in Fig. 4j, where two characteristic pressures $P_1$ and $P_2$ are clearly manifested in both parameters. The corresponding phase diagram for the $x$ = 0.67 sample is summarized in Fig. 4k, from which the pressure-induced metal-insulator-metal(superconductor) transitions are clearly visualized.

## Discussion

The interaction of localized electronic states and the conduction electron Fermi sea, enshrined in the Kondo framework, is widely recognized to be responsible for a wide spectrum of physical phenomena, including heavy fermion physics and more generally flat band physics [6,16]. Owing to the more compact nature of *f*-orbital shells relative to *d*-orbitals, heavy-fermion behaviors are generically observed in *f*-electron-based rare earth compounds. Conversely, *d*-orbitals are considerably extended in real space and as such, only a handful of *d*-electron-based heavy fermion systems have been reported to date. The most notable example of a *d*-electron heavy fermion is $LiV_2O_4$ [10,46,56]. In this vanadium pyrochlore-containing oxide, magnetic frustration due to its triangular structural motif, or proximity to a Mott insulating state, is widely accepted to play a crucial role in driving its heavy-fermion behaviors [57]. Moreover, heavy fermion and strange metal behaviors have recently been reported in the 3*d* transition metal kagome intermetallic $Ni_3In$ [14]. Its kagome lattice, composed of corner-sharing triangles, induces destructive quantum interference between lattice and orbital degrees of freedom, giving rise to hopping interference-driven localized moments. These frustration hopping-derived moments form a Kondo lattice and flattened electronic bands. In $Ni_3In$, pre-formed local magnetic moments are experimentally evidenced by its Curie-Weiss susceptibility at high temperatures; additionally, calculations of local and momentum-resolved magnetic susceptibility further support this local-moment scenario which originates from the partially filled flat bands [14]. Our $Rh_{17}S_{15}$ bears a close resemblance to $Ni_3In$, in that its magnetic susceptibility exhibits pronounced Curie-Weiss behavior at elevated temperature (Fig. S5) and its flat band electronic structure is also present near the Fermi energy. Once local moments nucleate, the low-energy physics of this *d*-electron system can be described by the interactions between these local moments and conduction electrons, in analogy to prototypal heavy fermions in Kondo *f*-lattice systems. Additionally, in terms of the Doniach phase diagram [58], a proximate magnetically order phase can be induced by tuning the hybridization constant $J$. This parameter is readily adjustable by external control parameters such as pressure, magnetic field or chemical doping, as the relevant energy scales are relatively small in heavy fermion compounds [59]. While this adjacent magnetic state has not yet been observed in $Rh_{17}S_{15}$ and Fe-doped samples, prospective experimental investigations exploring the broader phase space are of high interest to probe the underlying magnetic instabilities.

Likewise, flat band physics has recently been observed in $CaNi_2$, a three-dimensional metal with a nickel pyrochlore lattice [15]. Similarly, the flat bands are ascribed to the lattice geometry-enforced destructive quantum interference of electron motions, which suppresses electronic dispersion along all three dimensions. This is because the pyrochlore lattice can be viewed as the 3D counterpart of the 2D kagome lattice. Structurally, this 3D pyrochlore lattice arises from the extension of 2D kagome triangular plaquettes into the third dimension, an arrangement that enables destructive interference of electronic hopping along all spatial dimensions, thereby stabilizing the 3D flat bands. In distinct contrast to these triangular lattice materials, $Rh_{17}S_{15}$ possesses a highly symmetric cubic structure, disfavoring any geometric frustration.

As known, the ferromagnetism observed in magnetic susceptibility measurements can arise from either localized moments or itinerant ferromagnetism (also referred to as band ferromagnetism) [60]. Unlike many Rh-based compounds where Rh ions are *nonmagnetic*, the pronounced Curie-Weiss-like increase in susceptibility with decreasing temperature strongly indicates the formation of local magnetic moments in $Rh_{17}S_{15}$. These localized magnetic moments are likely closely related to the flat band illustrated in Fig. 1b. However, we note that the flat band alone cannot account for the emergence of localized moments as the calculations yield a zero net magnetic moment when electron-electron interactions are neglected. As depicted in Fig. S17 [61-63], a non-zero Rh magnetic moment only begins to emerge when a substantial Coulomb interaction $U$ is included ($U \geq 5$ eV). To reproduce a magnetic moment comparable to the experimental value, i.e., of the order of 0.1 $\mu_B$, a large Coulomb $U$ of ~6 eV is required (Fig. S5 and S18). This finding strongly supports the notion that the local magnetic moments in $Rh_{17}S_{15}$ indeed originate from strong *e-e* correlations. When correlation effects are incorporated, the electronic bands become further flattened and shifted closer to the Fermi level, leading to a drastically enhanced electronic density of states and significantly increased effective electron masses.

A more recent study has proposed a new paradigm for realizing heavy fermions in *d*-electron systems, that is, doping a Hund metal toward half-filling to induce a *frustrated* orbital-selective Mott transition [64]. It is well established that orbital-selective correlations strengthen significantly when approaching half-filling, favoring the onset of an orbital-selective Mott transition. However, the inter-orbital hopping suppresses such orbital-selective Mott *localization*; instead, it drives a *frustrated* orbital-selective Mott transition that serves as a source of the heavy-fermion behavior. In this framework, one part of *d*-orbitals tends to localize due to strong local correlations, resulting in significantly heavier electronic states than the remaining itinerant *d*-electrons. Importantly, these two parts remain hybridized, mimicking the interplay between *f*-electrons and conduction electrons in traditional heavy fermion systems. This scenario was originally proposed to explain heavy fermion behavior in Cr-doped $CsFe_2As_2$, known as a Hund's metal [64]. In undoped $CsFe_2As_2$, Fe ions adopt a nominal $d^{5.5}$ electronic configuration; these 5.5 electrons/Fe occupy the five *d*-orbitals following Hund's rule. Cr doping introduces additional holes and shifts the system toward half-filling (i.e, a $d^5$ configuration) and triggers a *frustrated* orbital-selective Mott transition. In our $Rh_{17}S_{15}$, Rh ions exhibit an approximate $d^7$ configuration. Fe doping can introduce holes and push the system toward half-filling. Within this scenario, however, the signature of Hundness and how the orbital-selective correlations induce frustrated Mottness and consequently drive the formation of heavy *d* electrons in $Rh_{17}S_{15}$ and Fe-doped samples remain unclear and call for further theoretical insights.

In the past, the field of strongly correlated electron systems has predominantly focused on materials with topologically trivial bands [16]. In contrast, the strongly correlated topological bands have only recently become a burgeoning research frontier. In this sense, this strongly correlated heavy *d*-electron $Rh_{17}S_{15}$ system could be further enriched by band topology in the form of novel gapless or gapped topological phases, as suggested by our first-principles calculations, paving a pathway to realizing topological-Kondo semimetals in *d*-electron-based materials. Consequently, our study could set a new paradigm for investigating the interplay between strong correlation effects, topological flat states and unconventional superconductivity.

.

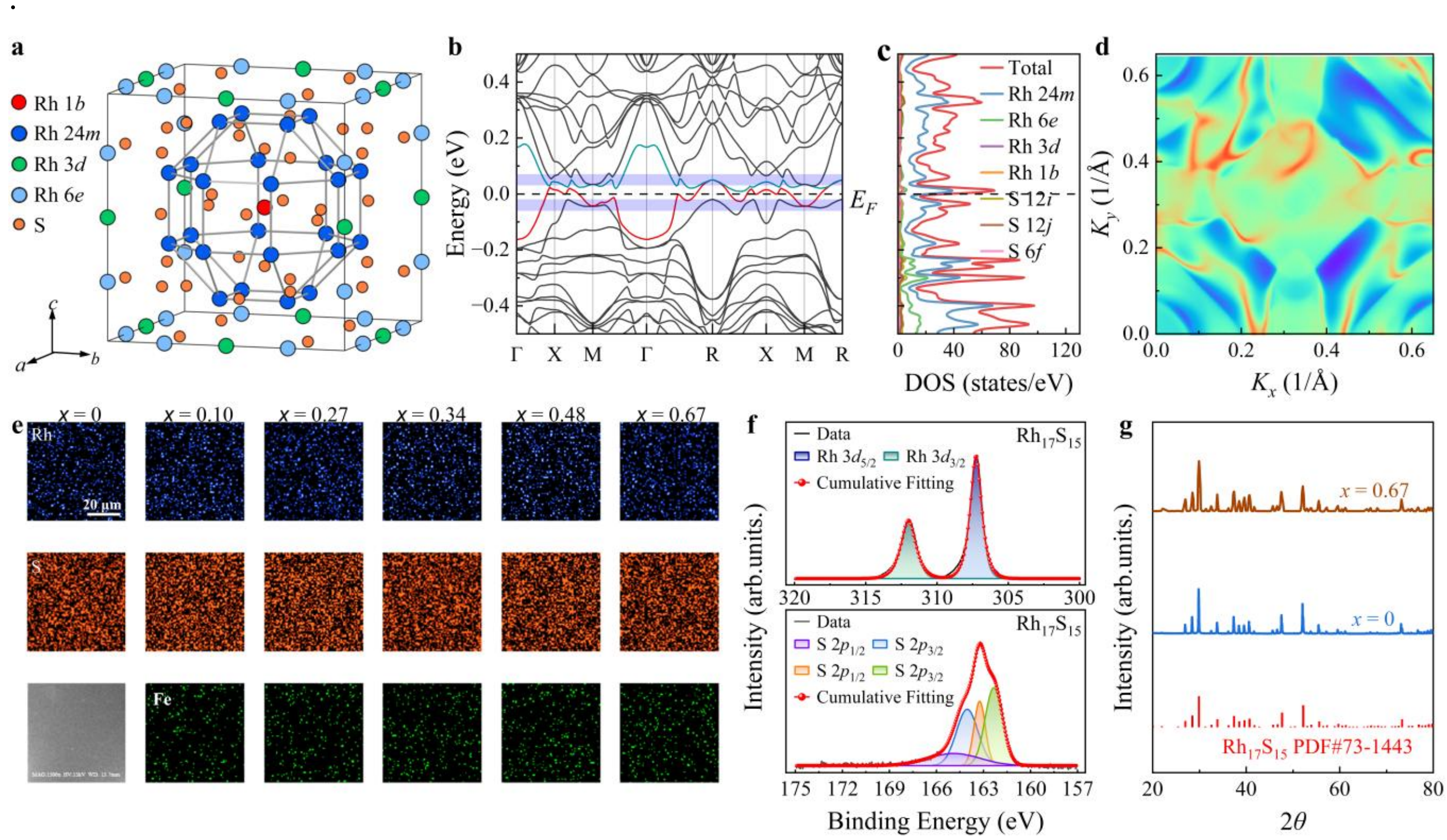

**Fig. 1 | Crystal structure and electronic structure.** **a** Crystal structure of $Rh_{17}S_{15}$. The unit cell has 4 types of Rh atoms with position symmetry 1*b*, 24*m*, 3*d*, and 6*e* and 3 types of sulfur atoms with position symmetry 12*j*, 12*i*, and 6*f*. For clarity, we use the same color for all sulfur atoms. **b** Electronic band structure showing two bands crossing the $E_F$, highlighted in red and green. Blue-shaded regions indicate flat band segments near $E_F$. **c** Total and projected DOS for Rh atoms at Wyckoff positions 24*m*, 6*e*, 3*d*, and 1*b*, and S atoms at positions 12*i*, 12*j*, and 6*f*. **d** Surface states on the (001) plane, where bright lines represent the surface state features. **e** The EDX mapping of Rh, S, and Fe for the as-grown single crystals revealing the homogeneous spatial distribution of the constituent elements. **f** The XPS spectra of Rh 3*d* and S 2*p* region, respectively. **g** Powder XRD patterns of the highest-doped samples and the undoped one. The $Rh_{17}S_{15}$-type structure (space group $Pm\bar{3}m$, PDF #73-1443) is included as a reference.

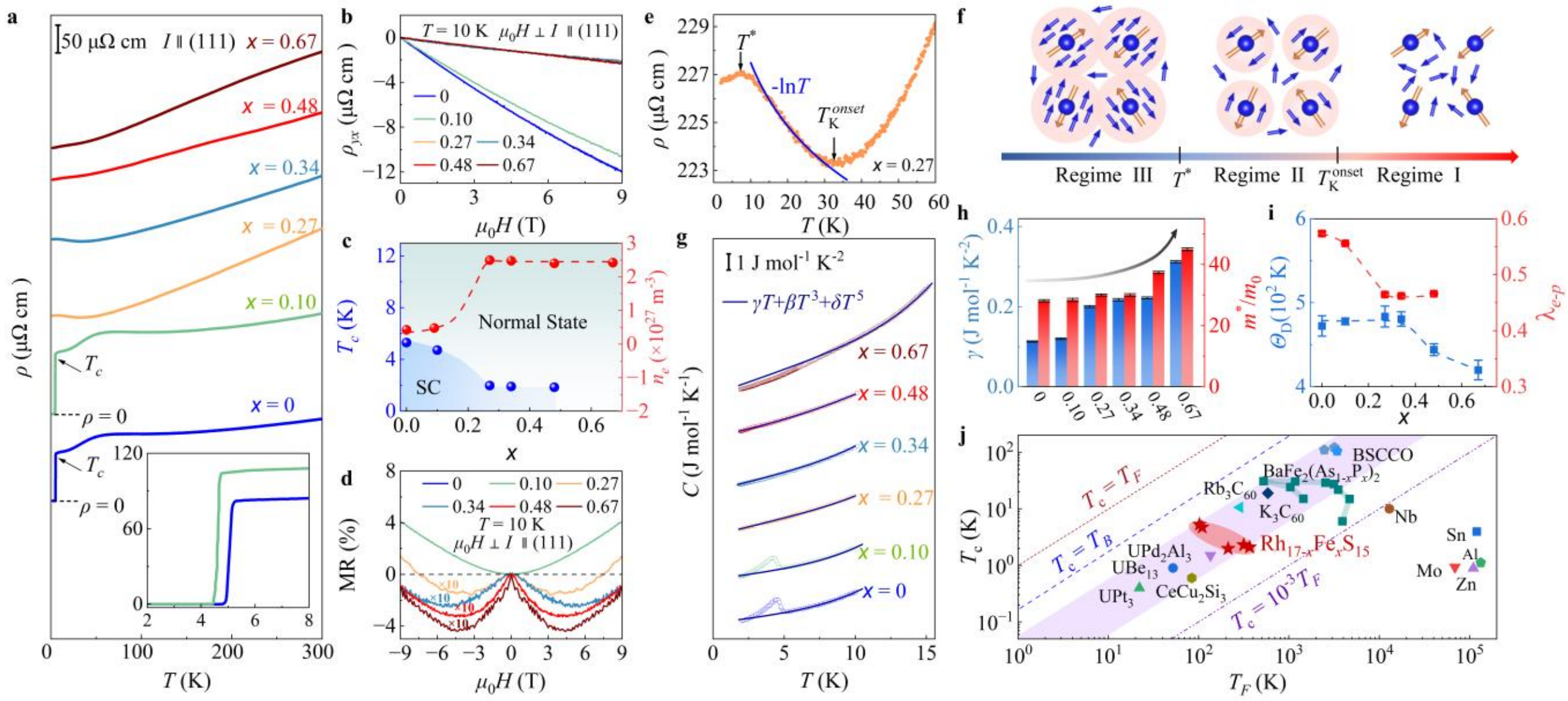

**Fig. 2 | Superconducting and transport properties of $Rh_{17-x}Fe_xS_{15}$ single crystals.** **a** Temperature dependence of resistivity for $Rh_{17-x}Fe_xS_{15}$ with $I \parallel (111)$, measured from 2 to 300 K. The inset shows the low-$T$ resistivity for $Rh_{17-x}Fe_xS_{15}$ ($x$ = 0 and 0.10). **b** Magnetic field dependence of Hall resistivity with $\mu_0 H \perp I \parallel (111)$ at 10 K. **c** Evolution of the superconducting transition temperature $T_c$, defined as the onset temperature of superconductivity, and the carrier concentration $n_e$ as a function of Fe content $x$. $T_c$ here denotes the onset temperature of superconductivity. **d** MR measured with $\mu_0 H \perp I \parallel (111)$ at 10 K for $Rh_{17-x}Fe_xS_{15}$. **e** Low-$T$ resistivity upturn for $Rh_{16.73}Fe_{0.27}S_{15}$. The blue solid line corresponds to $\rho \propto -ln(T)$ scaling. **f** Schematic diagrams of Kondo lattice screening. Regime Ⅰ ($T > T_K^{onset}$): non-screened regime; Regime Ⅱ ($T^* < T < T_K^{onset}$): incoherent Kondo regime; Regime Ⅲ ($T < T^*$): coherent Kondo regime. The hatched balls signify Kondo entanglement of localized moments (red) and conduction electrons (blue). **g** Temperature dependence of the specific heat for $Rh_{17-x}Fe_xS_{15}$ single crystal. The data are vertically offset for clarity. The lines represent the fits to $C = \gamma T + \beta T^3 + \delta T^5$ in the normal state. The detailed fitting range for each Fe-substitution level is provided in the Supplementary Information. **h** Sommerfeld coefficient $\gamma$ and effective mass $m^*/m_0$ for the $Rh_{17-x}Fe_xS_{15}$ single crystals. **i** Variation of the Debye temperature $\Theta$ and the electron-phonon coupling constant $\lambda_{e\text{-}p}$ with Fe concentration $x$. **j** Uemura plot of $T_c$ vs. $T_F$. $T_B$ here represents the Bose-Einstein condensation (BEC) temperature for an ideal 3D boson gas, given by $T_B = 0.176T_F$.

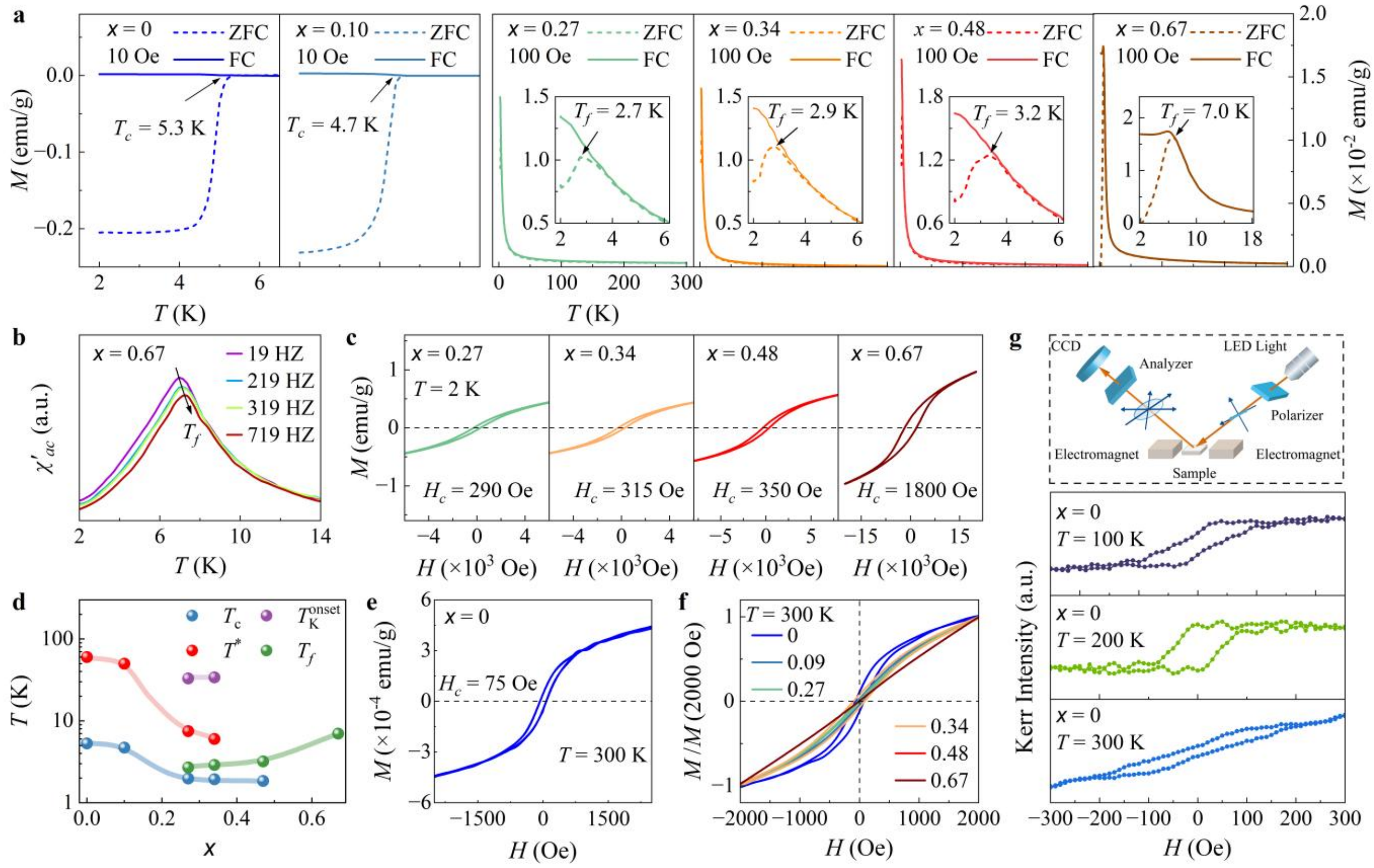

**Fig. 3 | Magnetic properties of $Rh_{17-x}Fe_xS_{15}$ single crystals. a** Temperature dependence of magnetization for $Rh_{17-x}Fe_xS_{15}$ measured in ZFC and FC modes under $\mu_0 H \perp (111)$. Insets show enlarged low-temperature regions. **b** Temperature dependence of in-phase component of the ac magnetic susceptibilities for $Rh_{16.33}Fe_{0.67}S_{15}$ in an ac magnetic field of 5 Oe at several fixed frequencies. **c** Magnetic field dependence of magnetization measured at 2 K for $Rh_{17-x}Fe_xS_{15}$ under $\mu_0 H \perp (111)$. f Temperature- composition phase diagram of $Rh_{17-x}Fe_xS_{15}$, including the superconducting transition temperature $T_c$, spin-glass transition temperature $T_f$, the onset of Kondo scattering $T_{\mathrm{K}}^{\mathrm{onset}}$ ,and the coherence temperature $T^*$. e Isothermal magnetization curves measured at 300 K for $Rh_{17}S_{15}$ under $\mu_0 H \perp (111)$. **f** Normalized isothermal magnetization curves for $Rh_{17-x}Fe_xS_{15}$ measured at 300 K under $\mu_0 H \perp (111)$. **g** Top: Schematic illustration of the MOKE measurement setup. Bottom: Magnetic field dependence of Kerr rotation hysteresis loops for $Rh_{17}S_{15}$ measured at 100, 200, and 300 K.

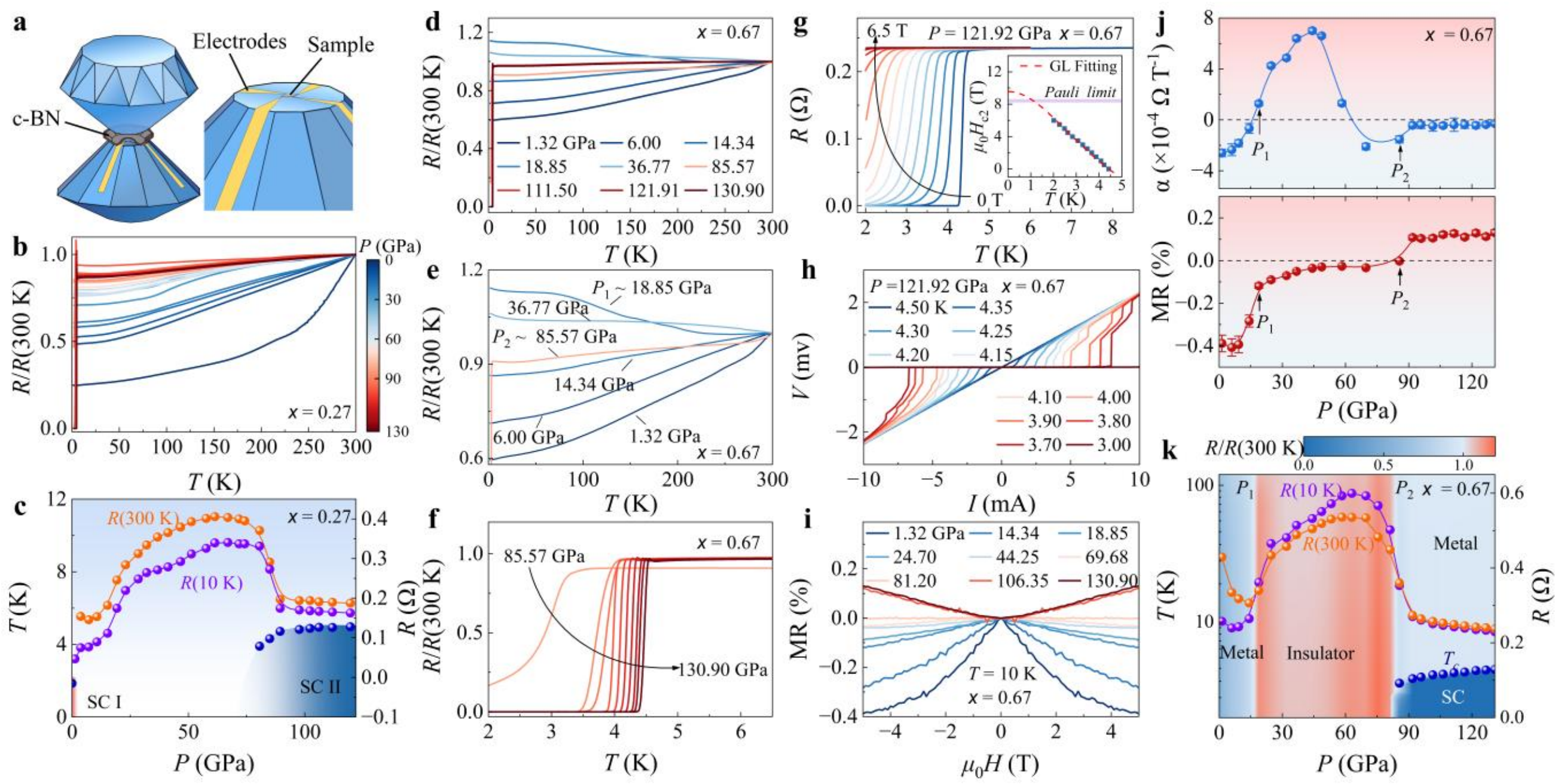

**Fig. 4 | Electrical transport properties and superconducting phase diagram of $Rh_{17-x}Fe_xS_{15}$ under pressure. a** Schematics for the high-pressure DAC setup. **b** Temperature dependence of normalized resistance $R/R$(300 K) for $x$ = 0.27 under various pressures, measured from 2 to 300 K. **c** Evolution of the superconducting transition temperature $T_c$, determined from the onset temperature of superconductivity, and the resistances at 10 K and 300 K as a function of pressure for $x$ = 0.27. The purple and brown symbols indicate resistance values at 10 K and 300 K, respectively. **d** Temperature dependence of normalized resistance $R/R$(300 K) for $x$ = 0.67 under various pressures, measured from 2 to 300 K. **e** Enlarged plot of $R/R_{300\ K}$ illustrating the pressure-dependent evolution of resistance behavior. $P_1$ and $P_2$ are identified as 18.85 GPa and 85.57 GPa, respectively. **f** Low-temperature normalized resistance under 85.75-130.90 GPa showing the emergence of superconductivity. **g** Temperature dependence of resistance under various magnetic fields at 121.92 GPa. The inset shows the upper critical field $\mu_0H_{c2}$ determined from the onset temperature of superconductivity at each field. The dashed line represents the fit to the GL formula. The horizontal line indicates the Pauli limit. **h** Characteristic *I-V* curves of $Rh_{16.33}Fe_{0.67}S_{15}$ under 121.92 GPa at various temperatures. **i** MR measured at 10 K under various pressures. **j** Pressure dependence of the Hall resistance slope $\alpha$ and the value of MR at 5 T and 10 K. **k** Color mapping of the normalized resistance $R/R_{300\ K}$ as a function of temperature and pressure for $Rh_{16.33}Fe_{0.67}S_{15}$. The red region represents the normal insulating phase, the blue region corresponds to the metallic phase, and the superconducting phase is shaded in dark blue. The orange symbols represent the $T_c$, which denotes the onset temperature of superconductivity. The purple and brown symbols indicate resistance values at 10 K and 300 K, respectively.